\documentclass[sigconf]{acmart}
\setcopyright{acmcopyright}
\copyrightyear{2022}
\acmYear{2022}
\acmDOI{10.1145/1122445.1122456}

\acmConference[RoPES 2022]{The 1st International Workshop on Recruiting Participants for Empirical Software Engineering}{May 17, 2022}{Pittsburgh, PA, USA}
\acmPrice{15.00}
\acmISBN{978-1-4503-XXXX-X/18/06}

\begin{document}

\title{Recruiting Software Engineers on Prolific}

\author{Daniel Russo}
\orcid{0000-0001-7253-101X}
\affiliation{%
  \institution{Department of Computer Science \\ Aalborg University}
  \city{Copenhagen}
  \country{Denmark}
}
\email{daniel.russo@cs.aau.dk}

\renewcommand{\shortauthors}{Daniel Russo}

\begin{abstract}
Recruiting participants for software engineering research has been a primary concern of the human factors community.
This is particularly true for quantitative investigations that require a minimum sample size not to be statistically underpowered.
Traditional data collection techniques, such as mailing lists, are highly doubtful due to self-selection biases. 
The introduction of crowdsourcing platforms allows researchers to select informants with the exact requirements foreseen by the study design, gather data in a concise time frame, compensate their work with fair hourly pay, and most importantly, have a  high degree of control over the entire data collection process.
This experience report discusses our experience conducting sample studies using Prolific, an academic crowdsourcing platform.
Topics discussed are the type of studies, selection processes, and power computation.
\end{abstract}





\maketitle

\section{Introduction}

Recruiting software professionals for empirical software engineering investigation is a typical barrier for several research questions, especially for independent researchers. 
Often, such hurdles are bypassed by using students as a proxy of software developers, jeopardizing the generalizability of the study's conclusion \cite{feldt2018four}.
Other options are to recruit developers through professional mailing lists or social media, which implies severe self-selection bias and ethical issues about developers' privacy \cite{gold2020ethical}.
On the other hand, scholars employed by big software organizations have direct access to a massive amount of developers (also above 3,500 \cite{ford2021tale}).
However, one company employs all those informants, limiting such studies' generalization.
Another typical issue is the sampling strategy. Most software engineering papers rely on convenience sampling \cite{baltes2020sampling}, representing a significant threat to validity. 

In other words, most performed research involving software engineers in our community, so far, suffers from severe limitations typically represented by the low generalizability of their conclusions. 

Sample studies have an enormous potential to guide us towards a more theory-driven discipline \cite{stol2015theory,stol2018abc}.
In such investigations, the relevant variables in a given population (i.e., of people or systems) or the relation between more characteristics are collected through informants.
The ultimate goal is to generalize the findings.
Using the taxonomy of Stol \& Fitzgerald \cite{stol2018abc}, while research conducted in natural settings, such as field studies, aim to understand a specific research phenomenon, contrived settings focus on causal relations primarily through experiments: neutral settings allow to investigate the world as it is.
With sample studies, we can isolate the typical noise of field study and focus on the characteristics and interactions of the identified variables.

Therefore, any investigation that can (i) operationalize research variables (e.g., software quality, size, number of commits, productivity, gender)\footnote{In this exposition, we will not consider construct validity issues.}, (ii) have some underlying hypotheses about the relationship of the researched variables, and (iii) have a sufficiently large enough population to ensure well-powered statistical analysis is an ideal candidate for a sample study research design.

Unfortunately, sample studies have three main barriers.
First, to ensure the representatives of the selected population, i.e., how to be sure that our informants are indeed software engineers?
Second, find enough data points to perform a statistically well-powered analysis.
Third, operationalize your variables to be sure that they reflect the investigated phenomenon.
Crowdsourcing platforms, such as Prolific\footnote{Prolific is an academic data collection platform: www.prolific.co.}, directly address the first two barriers.
Ensuring construct validity is a more complicated task, and we refer to Russo \& Stol for more practical advice \cite{russo2021pls}.

In the following, we will discuss how we addressed the selection process of human participants and dealt with power computations in some of our studies using Prolific  \cite{russo2020gender,russo2020predictors,russo2021daily,Russo2021Success,Russo2022NfC,van2021effect,cucolas2021impact,russo2021developers,russo2021understanding}.

\section{Selection Process}

Through Prolific, you have to channel your own survey.
By January 2022, Prolific has more than 150,000 active users.
We administered our surveys on Qualtrics\footnote{www.qualtrics.com.}.
Other scholars used free instruments, such as Google Forms \cite{Ralph2020pandemic}.
Ralph et al. explain in their paper why the use of Google Forms was not a good idea and, in hindsight, they would opt for Qualtrics.
On Prolific, the scholar can choose whenever the survey has to be completed using a smartphone, a tablet, or a desktop computer. 
We allowed only participants to fill the survey using a desktop computer to ensure the participants' focus.

The first selection process happens in Prolific, where participants self-selected themselves by providing several demographics. 
Although it is not possible to select software engineers directly, pre-screening helps to narrow down the relevant populations.
In our past studies, we selected ``have knowledge of software development techniques'',  ``have computer programming skills'', ``use technology at work (e.g., software) at least once a day'', and have an ``approval rate of at least 95\%''. The last criterion refers to the level of reliability of Prolific platform members in Prolific past surveys. 
By January 2022, the Prolific users with such characteristics are more than 15,500.
We do not suggest selecting per Industry since software engineers are employed in various sectors, not only software.
Similarly, we do not recommend opting for balanced gender distribution since it would bias the results, considering that women represent (unfortunately) a 20\%-30\% minority \cite{russo2020gender}.

Still, we can not be sure if these candidates are indeed software engineers.
They could also have a similar profession, e.g., control engineers.
Thus, proper screening is required through a pilot study.
Pilot studies are concise studies where specific software engineering competence is tested.
To do that, we used a one-minute time-boxed three-question multiple-choice survey.
Danilova et al. developed a list of the most compelling question to ask in such pilot studies; we strongly recommend using those \cite{danilova2021you}.

After the pilot study, the researcher can select only the candidates who responded correctly in the given time.
Once the cohort has been finalized and selected, they are then invited to take the complete survey.
Here, usual recommendations about running surveys apply (e.g., randomization).
In addition, we used a number (2-3) quality attention checks to ensure that informants are indeed focused on the survey.

\section{Power Analysis}

How many participants are needed in a study to minimize the probability of making Type II (false negative) errors?
Power assumptions should be at the very core of every sample study design.
To do that, G*Power is a very valuable tool\footnote{https://www.psychologie.hhu.de/arbeitsgruppen/allgemeine-psychologie-und-arbeitspsychologie/gpower.}.
There, based on the type of analysis, you can perform an \textit{a priori} or a sensitivity analysis. As a reviewer, you can also perform a \textit{post-hoc} analysis to assess whenever the sample size is large enough.
Russo \& Stol provide a short explanation about the use of G*Power for Structural Equation Modeling analyses. 

Although our primary concern is underpowered studies that lead to flawed theories, we are also worried about Type-I (false-positive) errors.
This typically happens when the analysis deals with a large number of variables.
In such cases, the alpha levels need to be adjusted accordingly \cite{russo2021developers}.
As a suggestion, we do not recommend using a Bonferroni correction since it is overly conservative. 
By modifying the alpha threshold, the test statistic (e.g., p-value, t-value) also needs to be modified.
For example, in Russo et al., we dealt with over 50 variables and performed a high number of tests \cite{russo2021developers}.
In that case, we considered significant only relations that had a p-value smaller than 0.003.

In most cases, computing the minimum sample size is a relatively straightforward task (when dealing with nested or higher-order data, things become more complex). Adjusting the alpha level is not.
Thus, our recommendation is to be guided by previous literature, especially methodologically more mature disciplines.
The software engineering community also underwent a significant challenge by developing the Empirical Standards \cite{ralph2020empirical}, which is an excellent starting point to start developing a proper sample study.

%\section{Conclusions}

\bibliographystyle{ACM-Reference-Format}
\bibliography{bib}

\end{document}